\documentclass[conference]{IEEEtran}
\IEEEoverridecommandlockouts
\usepackage[numbers,sort&compress]{natbib}

\usepackage{amsmath,amsfonts}
\usepackage{algorithm}
\usepackage{algpseudocodex}
\usepackage{booktabs}
\usepackage{graphicx}
\usepackage{textcomp}
\usepackage[dvipsnames]{xcolor}
\usepackage{fancyhdr}

\usepackage{hyperref}

\usepackage{tabularx}

\usepackage{subcaption}  % modern way to handle subfigures
\usepackage{caption}
\usepackage{wrapfig}

\usepackage{enumitem}
\setlist[itemize]{noitemsep, topsep=0pt, leftmargin=*}

\usepackage{xspace}

\usepackage{listings}
\usepackage{multirow}
\usepackage{threeparttable}

\usepackage[export]{adjustbox}
\usepackage{cleveref}

\newcommand{\ie}{\textit{i}.\textit{e}.,\xspace}

\usepackage{tikz}
\newcommand*\circled[1]{\tikz[baseline=(char.base)]{
		\node[shape=circle,fill,inner sep=1pt] (char) {\textcolor{white}{#1}};}}

\newcommand{\design}{VEGA\xspace}

\usepackage{setspace}
\setlist[itemize]{noitemsep, topsep=0pt, leftmargin=*}

\begin{document}
	
	\title{\design: A Scalable Software-Hardware Co-design for Genome Alignment using RVV Extension}
        
    \author{
	  \IEEEauthorblockN{
	    Weihong Xu\IEEEauthorrefmark{1}\textsuperscript{\textdaggerdbl},
	  	Shijie Xu\IEEEauthorrefmark{2},
	  	Qunyou Liu\IEEEauthorrefmark{2},
	  	Giovanni Ansaloni\IEEEauthorrefmark{2},
	  	Cheng Zhuo\IEEEauthorrefmark{1} and
	  	David Atienza\IEEEauthorrefmark{2}
	  }
	  \IEEEauthorblockA{
	  	\IEEEauthorrefmark{1}College of Integrated Circuits, Zhejiang University, Hangzhou, China\\
	  	\IEEEauthorrefmark{2}Embedded Systems Laboratory (ESL), École Polytechnique Fédérale de Lausanne (EPFL), Lausanne, Switzerland\\
      Email: \{wh.xu, czhuo\}@zju.edu.cn, \{shijie.xu, qunyou.liu, giovanni.ansaloni, david.atienza\}@epfl.ch
	  }
     \vspace{-0.5cm}
     \thanks{\textsuperscript{\textdaggerdbl}This work was carried out while Weihong Xu was a postdoctoral researcher with the Embedded Systems Laboratory (ESL), EPFL, Lausanne, Switzerland.}
    }
	\maketitle

    \thispagestyle{plain}
    \pagestyle{plain}

    \begin{abstract}
    Dynamic programming (DP)-based genome sequence analysis is fundamental to computational genomics, but it remains computationally intensive as genomic data volumes continue to outpace Moore's Law. Existing hardware accelerators often sacrifice programmability and scalability across diverse alignment algorithms. 
    In this work, we present \design, a scalable software–hardware co-design that accelerates genome alignment using RISC-V Vector extensions. \design introduces specialized vector instructions and three key innovations: a Forefront Stationary dataflow that minimizes off-chip memory access, inter-sequence parallelism that maximizes SIMD lane utilization, and custom ISA extensions that accelerate core alignment operations.
    Evaluations show that \design achieves up to $15.1\times$ and $9.7\times$ speedups for Edit Distance and Smith–Waterman algorithms over the Ara2-RVV vector baseline, sustaining $10.2\times$ and $6.7\times$ gains for 10kbp sequences. Post-synthesis results indicate only $1.0\%$ area overhead compared to a standard RISC-V core, with performance-per-area efficiency rivaling specialized ASICs. \design thus achieves a balanced advancement in efficiency, programmability, and scalability for genome sequence alignment.
    \end{abstract}

    \begin{IEEEkeywords}
    RISC-V, vector extension, genome alignment, vector processing, hardware accelerator.
    \end{IEEEkeywords}

	%%%%%%%%%%%%%%%%%%%%%%%%%%%%%%%%%%%%%%%%%%%%%%%%%%%%%%%%%%%%%%%%%%%%%%%%%%%%%%%%%%%%%%%%%%
	%%%%%%%%%%%%%%%%%%%%%%%%%%%%%%%%%%%%%%%%%%%%%%%%%%%%%%%%%%%%%%%%%%%%%%%%%%%%%%%%%%%%%%%%%%

\section{Introduction}
Genome sequencing underpins modern biology and medicine, enabling advances in personalized healthcare~\cite{ashley2016towards} and pathogen detection~\cite{byrd2014clinical}. The rise of next-generation sequencing (NGS) technologies~\cite{slatko2018overview} has produced petabyte-scale datasets with increasingly long reads, demanding efficient computational analysis.
Dynamic programming (DP) algorithms, such as Smith–Waterman–Gotoh~\cite{smith1981swg}, Needleman–Wunsch~\cite{needleman1970nw}, and edit distance~\cite{levenshtein1966edit_distance}, remain the gold standard for alignment accuracy but suffer from quadratic complexity. Optimizations such as banded alignment~\cite{chao1992aligning,suzuki2018introducing,shao2024bsalign} and wavefront alignment~\cite{marco2021wfa} improve efficiency by constraining computation along diagonals; yet, they still face significant runtime challenges on conventional CPUs. 

As genomic data growth outpaces Moore's Law, researchers are pursuing faster and more efficient hardware accelerators for increasingly complex alignment workloads. Recent advances using CPUs~\cite{pavon2024quetzal,doblas2023gmx,liu2023block,shao2024bsalign,suzuki2018introducing,li2018minimap2,marco2021wfa}, GPUs~\cite{zeni2024leveraging,ahmed2019gasal2,aguado2023wfa}, FPGAs~\cite{chen2021high,feng2025fpga}, ASICs~\cite{haghi2023wfasic,liao2018adaptively,gu2023gendp,turakhia2018darwin,cali2020genasm}, and near-memory processing (NMP)\cite{alonso2024bimsa,xu2023rapidx,mansouri2022genstore} have achieved notable speedups. However, most fixed-function accelerators target specific algorithms, and the diversity of DP-based alignment methods\cite{liu2023block,shao2024bsalign,suzuki2018introducing,li2018minimap2,marco2021wfa} still demands extensive hardware redesign for each implementation.

Recent accelerators~\cite{gu2023gendp,doblas2023gmx,haghi2023wfasic,pavon2024quetzal} aim to strike a balance between performance and programmability in genome alignment. GenDP~\cite{gu2023gendp} employs ASIC-based programmable units for multiple DP algorithms but requires algorithm-specific hardware. GMX~\cite{doblas2023gmx} extends the open RISC-V ISA~\cite{waterman2014riscv} with genome-specific instructions, offering flexibility but limited speedups. WFAsic~\cite{haghi2023wfasic} achieves high performance and energy efficiency for wavefront alignment on an RISC-V SoC, though it lacks generality. QUETZAL~\cite{pavon2024quetzal} uses ARM SVE2~\cite{sve2_spec} on the A64FX SoC~\cite{fujitsu_a64fx} to improve programmability, yet it remains constrained by the closed ARM ISA. Overall, ASIC-based designs offer superior speed but poor flexibility, while ISA extensions improve programmability with moderate performance gains.

The RISC-V Vector (RVV) 1.0 extension~\cite{rvv_spec} addresses key limitations of existing programmable accelerators. Its open ISA supports domain-specific vector instructions while preserving software compatibility—an advantage over closed ISAs. RVV's dynamic SIMD length further suits the irregular computation patterns of alignment algorithms.
Widely adopted in machine learning~\cite{wang2024speed}, high-performance computing~\cite{perotti2025spatz,minervini2023vitruvius}, and embedded systems~\cite{szymkowiak2024marian}, RVV remains largely unexplored for genomics~\cite{lopez2023risc}. It offers a unique opportunity to combine hardware efficiency with the algorithmic flexibility required for modern genomic applications.

By leveraging the RVV ISA, we present \design, an accelerator for genome alignment using specialized extensions. Built upon the open-source RVV processor~\cite{perotti2024ara2}, \design enhances vector processing for alignment workloads. Profiling (Section~\ref{subsec:motivation_profiling}) shows that existing RVV implementations suffer from inefficient memory access and instruction overhead. To overcome these bottlenecks, \design introduces targeted optimization schemes that eliminate key vector inefficiencies while preserving full programmability, achieving substantial performance gains across diverse alignment algorithms. 
The key contributions of this work are summarized as follows:
\begin{itemize}[leftmargin=*]
  % \item We propose the Forefront Stationary (FFS) dataflow that significantly reduces data movement cost. This approach transforms score vector loading operations into efficient on-chip movement, eliminating redundant memory access and reducing the total number of score data loading operations.
  % \item To address the poor SIMD lane utilization observed in existing RVV implementations, \design implements a parallelism scheme that processes multiple sequence pairs simultaneously. This approach maximizes vector lane utilization by dividing the SIMD vector lane into multiple forefront regions, enabling efficient processing of genomic data even with limited sequence parallelism.
  \item \design addresses RVV inefficiencies by integrating three dataflow optimizations: Forefront Stationary (FFS), Inter-sequence Parallelism (ISP), and a Fused Add–Min Operation. FFS minimizes data movement by transforming repeated score loading into on-chip transfers, while ISP divides the SIMD lane into multiple forefront regions to process several sequence pairs concurrently, maximizing lane utilization. The fused add–min instruction further reduces arithmetic overhead by combining consecutive operations. Together, these optimizations enhance throughput, efficiency, and programmability.
  \item \design extends the RVV ISA with domain-specific instructions tailored for genomic alignment. These extensions introduce only 1.0\% area overhead while significantly improving computational efficiency for genomic workloads.
  \item Our evaluation results demonstrate that \design achieves $15.1\times$ and $9.7\times$ speedup for the Edit Distance and Smith-Waterman algorithms over the Ara2-RVV baseline for short reads while maintaining substantial advantages ($10.2\times$ and $6.7\times$ speedups) for 10 kbp sequences. This scalability makes \design effective across diverse sequence lengths in modern genomic applications.
\end{itemize}

%%%%%%%%%%%%%%%%%%%%%%%%%%%%%%%%%%%%%%%%%%%%%%%%%%%%%%%%%%%%%%%%%%%%%%%%%%%%%%%%%%%%%%%%%%
%%%%%%%%%%%%%%%%%%%%%%%%%%%%%%%%%%%%%%%%%%%%%%%%%%%%%%%%%%%%%%%%%%%%%%%%%%%%%%%%%%%%%%%%%%
\section{Background and Motivation}

%%%%%%%%%%%%%%%%%%%%%%%%%%%%%%%%%%%%%%%%%%%%%%%%%%%%%%%%%%%%%%%%%%%%%%%%%%%%%%%%%%%%%%%%%%
\subsection{Dynamic Programming-based Genome Alignment}

% Genome sequence alignment is the fundamental computational problem in bioinformatics to comparative analysis for biological sequences, such as DNA, RNA, and protein sequences. Dynamic programming (DP) algorithms serve as the cornerstone for optimal sequence alignment. These algorithms construct a two-dimensional matrix where each cell represents the optimal score for aligning sequence prefixes up to that position. 
% DP alignment algorithms include several variants. For example, Smith-Waterman-Gotoh (SWG) algorithm~\cite{smith1981swg} implements optimal local sequence alignment using affine gap penalties that distinguish between gap openings and extensions. Needleman-Wunsch (NW) algorithm~\cite{needleman1970nw} employs similar principles for global alignment, while edit distance~\cite{levenshtein1966edit_distance} provides a simplified alignment metric representing the minimum operations required to transform one sequence into another. 

Genome sequence alignment is a core bioinformatics problem for comparing DNA, RNA, and protein sequences. Dynamic programming (DP) algorithms form their foundation, computing a matrix where each cell stores the optimal alignment score for sequence prefixes. Commonly used variants include Smith–Waterman–Gotoh (SWG)\cite{smith1981swg} for local alignment with affine gaps, Needleman–Wunsch (NW)\cite{needleman1970nw} for global alignment, and edit distance~\cite{levenshtein1966edit_distance} for minimal edit transformations.

\begin{figure}[ht]
  \centering
  \includegraphics[width=0.95\linewidth]{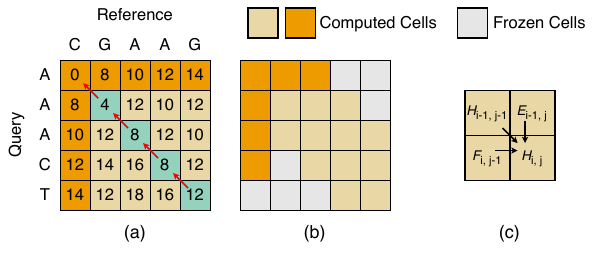}
  \caption{(a) Score computation and traceback, (b) banded alignment, and (c) data dependency graph of DP alignment.}
  \label{fig:dp_example}
\end{figure}

Fig.~\ref{fig:dp_example}-(a) illustrates a SWG example of the DP score matrix between the reference sequence ``CGAA'' and the query sequence ``AACA''. Given two sequences, \(Q = q_1 q_2 \dots q_m\) and \(R = r_1 r_2 \dots r_n\), a match/mismatch scoring function \(s(\cdot,\cdot)\), a gap-opening penalty \(o\), and a gap-extension penalty \(e\), the recurrence relations for the affine-gap SWG algorithm are as follows:
\begin{equation}
% \footnotesize
\begin{gathered}
% \footnotesize
  E_{i,j} = \min
  \begin{cases}
      H_{i-1,j} + o + e \\
      E_{i-1,j} + e
  \end{cases}
  F_{i,j} = \min
  \begin{cases}
      H_{i,j-1} + o +e \\
      F_{i,j-1} + e
  \end{cases}       
  \\
  % \centering
% \scriptsize
  H_{i,j} = \min
  \begin{cases}
      E_{i,j} \\
      F_{i,j} \\
      H_{i-1,j-1} + s(q_i,r_j) 
  \end{cases}              
\end{gathered}
\end{equation}
where \(H(i,j)\) is the optimal alignment score ending at positions \(i,j\). \(E(i,j)\) and \(F(i,j)\) are scores for alignments ending with a gap in the query (\(Q\)) or reference (\(R\)) sequences, respectively. \(s(q_i,r_j)\) is the substitution scoring function, typically adopting a match/mismatch scheme for DNA/RNA sequences. The DP-based algorithms first fill the matrix based on these recursive equations. After computing all cell values, the optimal alignment is obtained by performing a traceback from the highest-scoring cell (for local alignments) or from the last cell (for global alignments), reconstructing the path of optimal scores to determine the alignment path.

Although DP-based algorithms guarantee optimal solutions, their quadratic \(\mathcal{O}(mn)\) computational complexity and memory footprint are major bottlenecks for modern genomics workloads. To improve the computation efficiency, various optimization techniques have been developed. Fig.~\ref{fig:dp_example}-(b) depicts the banded alignment~\cite{chao1992aligning}, which constrains computation to an anti-diagonal region in the DP matrix, reducing computational complexity for similar sequences from $\mathcal{O}(mn)$ to $\mathcal{O}(\text{BW}\cdot n)$, where $\text{BW}$ denotes the band width.

% The other optimized DP alignment algorithm is Wavefront Alignment Algorithm (WFA)~\cite{marco2021wfa}. Instead of computing the entire DP matrix, WFA explores the alignment space along diagonals of matching bases (wavefronts), extending these diagonals incrementally to construct the optimal alignment path. WFA with $\mathcal{O}(s^2)$ complexity, shows promising speedup when processing short and accurate reads because the yield score $s$ is small. However, several prior works~\cite{liu2023block,shao2024bsalign} observe that WFA's performance degrades substantially for high-error sequences due to the large $s$ value. Therefore, it is challenging to implement efficiently for error-prone sequences.

The Wavefront Alignment Algorithm (WFA)~\cite{marco2021wfa} optimizes DP by expanding diagonals of matching bases (wavefronts) to build the alignment path, achieving $\mathcal{O}(s^2)$ complexity. It accelerates short, accurate reads with small yield scores $s$, but prior studies\cite{liu2023block,shao2024bsalign} report significant slowdowns on high-error sequences, making efficient implementation challenging for error-prone data.

%%%%%%%%%%%%%%%%%%%%%%%%%%%%%%%%%%%%%%%%%%%%%%%%%%%%%%%%%%%%%%%%%%%%%%%%%%%%%%%%%%%%%%%%%%
\subsection{Vectorized Genome Alignment}

\begin{figure}[t]
  \centering
  \includegraphics[width=0.95\linewidth]{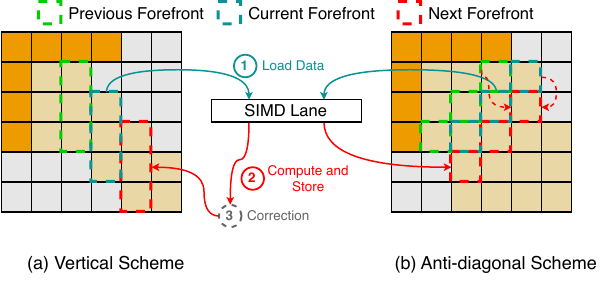}
  \caption{Comparison of two vectorized dataflows for banded DP alignment: (a) vertical and (b) anti-diagonal schemes.}
  \label{fig:vec_dp}
\end{figure}

In this work, we focus on banded DP alignment\cite{suzuki2018introducing,chao1992aligning,liu2023block,shao2024bsalign,li2018minimap2} for its scalability and parallelism in long-read sequencing.
Fig.~\ref{fig:vec_dp} compares two vectorized banded alignments (vertical and anti-diagonal schemes) commonly used in modern aligners~\cite{liu2023block,shao2024bsalign,suzuki2018introducing,li2018minimap2} and accelerators~\cite{pavon2024quetzal,gu2023gendp,feng2025fpga,liao2018adaptively,zeni2024leveraging}. Both load data to SIMD lanes and update scores but differ in forefront organization. 
\textbf{The vertical scheme} is simple to map in memory but suffers from vertical dependencies ($E_{i,j}$ depends on $E_{i-1,j}$). Prior prefix-scan~\cite{liu2023block} and F-loop~\cite{farrar2007striped,shao2024bsalign} optimizations only partly mitigate this issue. \textbf{The anti-diagonal scheme} removes such dependencies, enabling full parallelism~\cite{pavon2024quetzal,feng2025fpga,zeni2024leveraging}, though it requires string reversal~\cite{liu2023block,suzuki2018introducing} and modified memory layouts.
\design adopts the anti-diagonal approach to exploit RVV's high parallelism, introducing optimized dataflow and memory strategies to overcome these challenges.

%%%%%%%%%%%%%%%%%%%%%%%%%%%%%%%%%%%%%%%%%%%%%%%%%%%%%%%%%%%%%%%%%%%%%%%%%%%%%%%%%%%%%%%%%%
\subsection{RISC-V Vector Extension}

The RISC-V Vector (RVV) extension~\cite{rvv_spec} offers a flexible, scalable framework for vector processing. Unlike fixed-width SIMD architectures such as x86 AVX or ARM NEON~\cite{arm_neon}, RVV 1.0 introduces a length-agnostic model that ensures software portability across hardware with different vector lengths. Its configurable \texttt{VLEN} and runtime-tunable parameters via \texttt{vsetvli} enable efficient adaptation to diverse workloads.
Although RVV is widely adopted in machine learning and HPC~\cite{wang2024speed,minervini2023vitruvius,perotti2025spatz,szymkowiak2024marian}, its potential for genomic sequence alignment remains unexplored. Existing accelerators like GMX~\cite{doblas2023gmx} and WFAsic~\cite{haghi2023wfasic} rely on RISC-V scalar instructions, overlooking RVV’s parallelism. This work bridges that gap with an RVV-based acceleration framework incorporating specialized vector extensions for efficient genomic computation.

% \begin{listing}[t]
% \footnotesize
% \begin{minted}[linenos]{c}
\begin{lstlisting}[
  language=Python,
  basicstyle=\ttfamily\footnotesize,
  caption={Pseudocode of SIMD-vectorized banded alignment.},
  label={code:vec_banded_dp},
  frame=single
]
n_forefront = m+n+1;
for(h=2;h<n_forefront;++h) {
  for(v=1;v<bandwidth+1;v+=16) {
    /* Compute New E Forefront */
    v_prev_E = load_score(E);
    v_prev_H = load_score(H);
    v_prev_E = saddu_u16(v_prev_E, e);
    v_prev_H = saddu_u16(v_prev_H, o+e);
    v_E = min_u16(v_prev_E, v_prev_H);
    store_score(E, v_E); 
    /* Compute New F Forefront */
    ...
    /* Load Ref. and Query String */
    load_str_rev(str_ref, v_ref); //Load and Reverse
    load_str(str_query, v_query);
    v_score = blendv(mismatch, match, cmpeq(vquery, vref))    
    /* Compute New H Forefront */
    v_prev_H = load_score(H);
    v_H = saddu_u16(v_prev_H, v_score);
    v_H = min_u16(v_E, v_H);
    v_H = min_u16(v_F, v_H);
    store_score(H, v_H); 
  }
}
\end{lstlisting}
% \end{minted}
% \caption{Pseudocode of SIMD-vectorized banded alignment.}
% \label{code:vec_banded_dp}
% \end{listing}

%%%%%%%%%%%%%%%%%%%%%%%%%%%%%%%%%%%%%%%%%%%%%%%%%%%%%%%%%%%%%%%%%%%%%%%%%%%%%%%%%%%%%%%%%%
\subsection{Motivational Analysis}\label{subsec:motivation_profiling}
To evaluate banded DP alignment with anti-diagonal dataflow, we implement and profile two vectorized C kernels (Listing~\ref{code:vec_banded_dp}) using AVX2 and RVV, respectively. Experiments are conducted on a 16-thread AMD Ryzen 7 9700X~\cite{amd_ryzen_7_9700x} and the open-source Ara2 RVV 1.0 processor~\cite{perotti2024ara2}. Additional setup details are provided in Section~\ref{sec:eval}. We summarize the profiling results into following insights:

\begin{figure}[t]
  \centering
  \begin{subfigure}[b]{0.45\linewidth}
      \includegraphics[width=\linewidth]{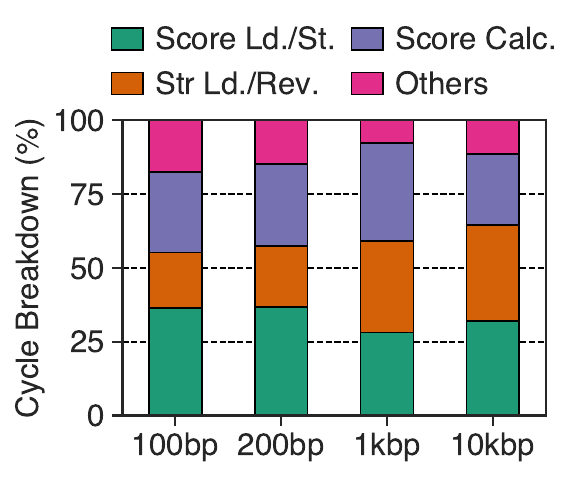}
      \caption{Cycle breakdown of AVX2 kernels on CPU}
      \label{subfig:avx2}
  \end{subfigure}   
  \hfil
  \begin{subfigure}[b]{0.45\linewidth}
      \centering
      \resizebox{\linewidth}{!}{
          \scriptsize
          \begin{tabularx}{\linewidth}{c|c}
              \hline
              \textbf{ISA} & \textbf{SIMD Width} \\
              \hline
              AVX2~\cite{intel_avx2_2022} & 256b \\
              ARM SVE2~\cite{sve2_spec} & 128-2048b \\
              RISC-V V 1.0~\cite{rvv_spec} & Flexible \\
              \hline
              \hline
              \textbf{Algorithm} & \textbf{Band Width} \\
              \hline
              ABEA~\cite{feng2025fpga} & 100 \\
              BSAlign~\cite{shao2024bsalign} & 32-1024 \\
              Block Aligner~\cite{liu2023block} & 32-128 \\
              libgaba~\cite{suzuki2018introducing} & 16-256 \\
              minimap2~\cite{li2018minimap2} & 500-1000 \\
              \hline
          \end{tabularx}
      }
      \caption{Existing ISA and banded algorithms}
      \label{tab:bw_sota}
  \end{subfigure}   
  \begin{subfigure}[b]{0.45\linewidth}
      \includegraphics[width=\linewidth]{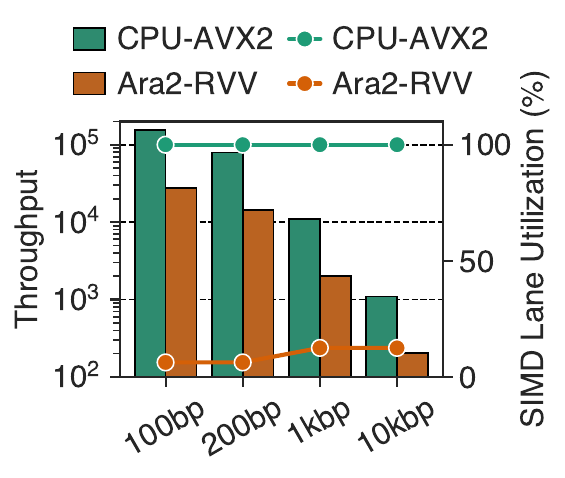}
      \caption{Single-core performance of AVX2 and RVV kernels}
      \label{subfig:perf_avx2_rvv}
  \end{subfigure}
  \hfil
  \begin{subfigure}[b]{0.45\linewidth}
      \includegraphics[width=\linewidth]{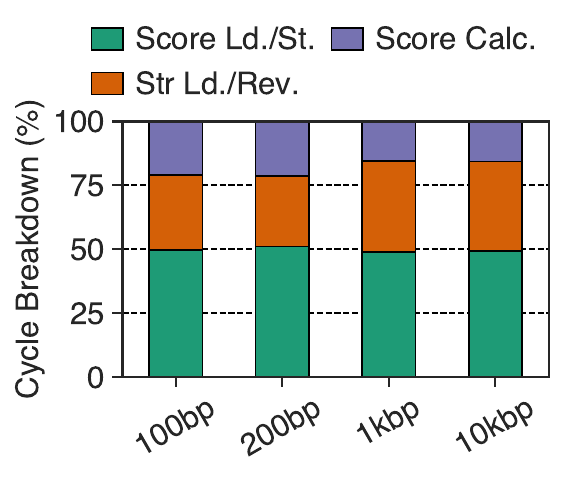}
      \caption{Cycle breakdown of RVV kernels on Ara2~\cite{perotti2024ara2}}
      \label{subfig:cycle_breakdown}
  \end{subfigure}   
  \caption{Profiling results of vectorized banded SWG.}
  \label{fig:profiling}
\end{figure}

% \noindent\textbf{Challenge 1: Instruction and IO overhead.}
% The cycle breakdown in Fig.~\mbox{\ref{fig:profiling}}-(a) suggests that banded DP alignment on x86 CPU is dominated by memory operations with over 50\% of cycles being dedicated to data loading or storing. This is due to the data movement for score/string loading and storing. The alignment computation is frequently stalled by the memory operations, degradating overall arithmetic unit utilization. As a result, only $\approx25\%$ of cycles are used for score calculation.
\noindent\textbf{Challenge 1: Instruction and I/O overhead.}
As shown in Fig.~\ref{fig:profiling}-(a), over 50\% of CPU cycles are spent on memory operations for score and string loading/storing, frequently stalling computation and reducing arithmetic unit utilization, leaving only about 25\% of cycles for actual score calculation.

% \noindent\textbf{Challenge 2: Limited SIMD width.} The limited SIMD width leads to excessive instruction and data movement overhead. As shown in Fig.~\ref{fig:profiling}-(b), there exists a huge gap between the band width of existing banded alignment algorithms and the available data parallelism in SIMD ISAs. The AVX2 and ARM SVE2 ISAs are not capable of updating the forefront score vector in one iteration. State-of-the-art banded alignment algorithms~\cite{shao2024bsalign,liu2023block,suzuki2018introducing,feng2025fpga,li2018minimap2} adopt a band width between 16 to 1024. For AVX2, the inner loop in Line 3 of Listing~\ref{code:vec_banded_dp} needs to be executed for $\frac{16\cdot \text{BW}}{256}$ iterations to update the forefront score vector in 16b integer, which further exacerbates the instruction overhead and limits the alignment throughput. Therefore, this necessitates a more scalable ISA with wider SIMD width.

\noindent\textbf{Challenge 2: Limited SIMD width.}
As in Fig.~\ref{fig:profiling}-(b), the narrow SIMD width of AVX2 and ARM SVE2 creates a large mismatch with the 16–1024 band widths used in modern aligners\cite{shao2024bsalign,liu2023block,suzuki2018introducing,feng2025fpga,li2018minimap2}. For a 256b-wide SIMD format such as AVX2, the inner loop in Listing~\ref{code:vec_banded_dp} must run $\frac{16\cdot \text{BW}}{256}$ times to update a 16b forefront vector, causing excessive instruction overhead and limiting throughput. This necessitates a more scalable and wider SIMD ISA.

% \noindent\textbf{Challenge 3: RVV inefficiencies.}
% To address instruction overhead and narrow SIMD width limitations, we exploit scalable RVV with its flexible SIMD width that reduces instruction fetch/issue requirements, \eg a RVV processor with \texttt{VLEN}=4096b needs only one instruction for \texttt{add} or \texttt{min} operation with maximum $\frac{\texttt{VLEN}}{16}=256$ elements. We rewrite the banded DP alignment kernels using RVV 1.0 ISA~\cite{rvv_spec} and implemented them on the open-source Ara2 processor~\cite{perotti2024ara2} with \texttt{VLEN}=4096b. However, Fig.~\ref{fig:profiling}-(c) shows Ara2-RVV achieves significantly lower single-core throughput than x86 AVX2 despite nearly 100\% lane utilization. This stems from two factors: poor SIMD lane utilization (BW=16 or 32 saturates only 12.5-25\% of 4096b \texttt{VLEN}) and memory operation bottlenecks. As shown in Fig.~\ref{fig:profiling}-(d), score loading/storing consumes 50-55\% of cycles, string loading/reverse operations consume 25-30\%, while actual score calculation accounts for merely 15-20\%. The costly string operations result from slow \texttt{vrgather.vv} instructions in RVV, which use sequential FIFO-based queues requiring multiple cycles to gather data from different SRAM banks across Ara2 and other RVV processors~\cite{minervini2023vitruvius,chipsalliancet1_2025,ucb-barsaturn-vectors_2025}.

\noindent\textbf{Challenge 3: RVV inefficiencies.}
To overcome instruction overhead and limited SIMD width, we leverage RVV's scalable SIMD design, where a 4096b \texttt{VLEN} executes a single \texttt{add} or \texttt{min} operation on 256 elements. We implement RVV 1.0 kernels~\cite{rvv_spec} on the Ara2 processor~\cite{perotti2024ara2}, but Fig.~\ref{fig:profiling}-(c) shows Ara2-RVV underperforms x86 AVX2 despite near-full lane utilization. The bottlenecks arise from low SIMD lane usage (BW=16–32 uses only 12.5–25\% of 4096b) and heavy memory overhead (50–55\% of cycles for score I/O, 25–30\% for string handling, and only 15–20\% for computation, as shown in Fig.~\ref{fig:profiling}-(d)). 

\section{Proposed \design Accelerator}\label{sec:framework}

\subsection{Overview}

\begin{figure}[t]
  \centering
  \includegraphics[width=0.9\linewidth]{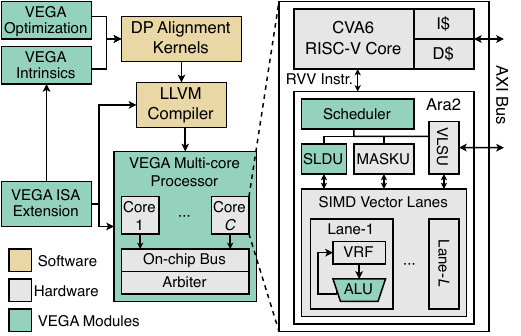}
  \caption{Overall architecture of \design framework.}
  \label{fig:arch_proposed}
\end{figure}

Fig.~\ref{fig:arch_proposed} illustrates the proposed \design software–hardware co-design for genome alignment acceleration. \design integrates RISC-V scalar and vector extensions with customized software, efficiently implementing the proposed RVV-based extensions (Section\ref{subsec:isa_ext}) while maintaining full RISC-V compatibility. It provides a scalable multi-core solution for diverse alignment algorithms, achieving both programmability and high performance through the following two key design strategies:

\noindent\textbf{\design Software} includes genome-optimized dataflows and intrinsics to exploit RVV parallelism with minimal code changes fully. DP alignment kernels are optimized using three techniques in Section~\ref{subsec:proposed_dataflow_opt}: forefront stationary (FFS) dataflow, inter-sequence parallelism, and fused add–min operations. Then the kernels are compiled with the LLVM toolchain supporting \design ISA extensions and executed on the \design multi-core processor.

\noindent\textbf{\design Hardware} extends the CVA6 RISC-V core~\cite{zaruba2019cva6} with vector capabilities built on Ara2~\cite{perotti2024ara2}, a state-of-the-art RVV 1.0 processor~\cite{rvv_spec}. Key components include the VLSU for memory transfers, SLDU for vector reshuffling, MASKU for masking, and $L$ SIMD lanes (configurable up to 4096b \texttt{VLEN}), each with local VRFs and specialized ALUs. 
To overcome RVV’s limitations in genomic alignment, we enhance Ara2’s scheduler, SLDU, and ALU. The scheduler is extended to support \design ISA control, the SLDU optimized for fast string loading and reversal, and the ALU upgraded for fused add–min operations. These targeted modifications reduce instruction and data redundancy, maximizing RVV parallelism for DP alignment. Microarchitectural details are provided in Section~\ref{subsec:isa_ext}.

\begin{figure}[t]
  \centering
  \includegraphics[width=0.95\linewidth]{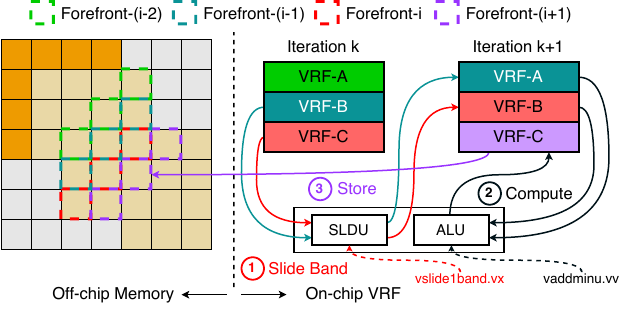}
  \caption{Forefront Stationary (FFS) dataflow in \design.}
  \label{fig:ffs_dataflow}
\end{figure}

%%%%%%%%%%%%%%%%%%%%%%%%%%%%%%%%%%%%%%%%%%%%%%%%%%%%%%%%%%%%%%%%%%%%%%%%%%%%%%%%%%%%%%%%%%
\subsection{Dataflow Optimization Strategies}\label{subsec:proposed_dataflow_opt}

\noindent\textbf{Forefront Stationary Dataflow.}\label{subsec:ffs_dataflow}
Section~\ref{subsec:motivation_profiling} demonstrates that the main bottleneck of DP alignment on RVV hardware is the high data-movement cost from repetitive score vector loading. Banded DP alignment, however, exhibits a data reuse opportunity along the anti-diagonal direction~\cite{pavon2024quetzal,suzuki2018introducing}, since Fig.~\ref{fig:ffs_dataflow} shows that each new forefront depends only on the previous two. Prior CPU implementations~\cite{pavon2024quetzal,suzuki2018introducing} fail to exploit this locality because the limited SIMD width requires data transfers to either cache or off-chip memory for large bands.

Leveraging RVV's wide vector register file (VRF), \design introduces the Forefront Stationary (FFS) dataflow in Fig.~\ref{fig:ffs_dataflow} to minimize data movement costs by keeping intermediate results stationary in VRFs. 
The FFS dataflow includes three steps: 
\circled{1} the previous forefront $i-1$ in VRF-B and the current forefront $i$ stored in VRF-C are first shifted based on the moving direction of the next forefront $i+1$. This is realized using the sliding operation performed in the SLDU. The vector data of previous and current forefronts do not go off-chip. Instead, they are stored in new VRFs, \ie VRF-A and VRF-B, for the following computation. 
\circled{2} the shifted forefronts are then fetched and loaded into the SIMD ALU to update the next forefront $i+1$, which is written back to VRF-C. 
\circled{3} The vector of updated forefront $i+1$ is stored back to off-chip memory since it is required by the traceback (TB) process. For each iteration, \circled{1} to \circled{3} are repeated. 
FFS drastically reduces external bandwidth usage, enabling efficient pipelining and high SIMD utilization. Compared with the naive kernel, score loading drops from $N_{\text{forefront}}$ to just one per alignment.

\begin{figure}[t]
  \centering
  \includegraphics[width=0.95\linewidth]{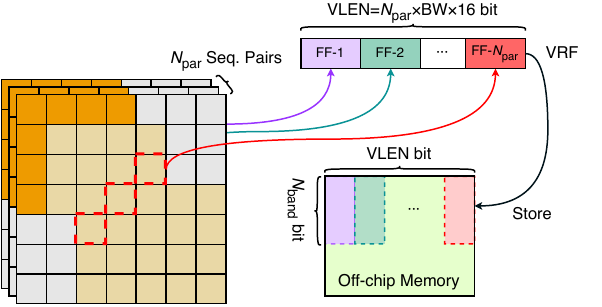}
  \caption{Inter-sequence parallelism (ISP) and its memory layout.}
  \label{fig:inter_seq_par}
\end{figure}

%%%%%%%%%%%%%%%%%%%%%%%%%%%%%%%%%%%%%%%%%%%%%%%%%%%%%%%%%%%%%%%%%%%%%%%%%%%%%%%%%%%%%%%%%%
\noindent\textbf{Inter-sequence Parallelism.}\label{subsec:inter_seq_par}
The previous FFS dataflow enables forefront-level parallelism but cannot fully utilize RVV's wide SIMD lanes. To address this, \design introduces inter-sequence parallelism (ISP) in Fig.~\ref{fig:inter_seq_par}, which complements FFS by processing $N_{\text{par}}$ sequence pairs concurrently with each occupying part of the \texttt{VLEN}-bit SIMD lane. The relationship between vector length and band width is $\texttt{VLEN}=N_{\text{par}}\times \text{BW}\times16$, allowing up to $N_{\text{par}}=\frac{\texttt{VLEN}}{\text{BW}\times16}$ sequences with 16-b precision to be processed simultaneously.
Unlike parasail~\cite{daily_parasail_2016} and swipe~\cite{rognes_faster_2011}, which assign independent alignments to separate SIMD units, \design's ISP preserves FFS data locality and minimizes off-chip bandwidth demand. Filling the SIMD lanes with multiple forefronts reduces instruction overhead—each update for $N_{\text{par}}$ forefronts requires only one instruction—and boosts throughput, even for narrow-band alignments.

The ISP dataflow also requires a coalesced memory layout in Fig.~\ref{fig:inter_seq_par}. DP matrices are stored in a continuous $\texttt{VLEN}\times N_{\text{forefront}}$ array, where each row holds $N_{\text{par}}$ forefronts. During alignment, (1) the first $BW$ rows are initialized, (2) the first two forefronts are loaded into VRFs for computation, and (3) only updated forefronts are written back. On-chip handling of band shifts and padding ensures aligned \texttt{VLEN}-bit transfers, maximizing memory bandwidth and avoiding unaligned-access penalties.

\noindent\textbf{Fused Add–Min Operation.}
DP alignment relies heavily on addition and minimum operations, which dominate instruction counts. To improve efficiency, we introduce a fused add–min operation that combines consecutive add and min instructions (\texttt{vaddminu.vv}/\texttt{vaddminu.vx}) into a single instruction, reducing instruction overhead. ISA and implementation details are provided in Section~\ref{subsec:isa_ext}.

% \noindent\textbf{Fused Add-Min Operation.}
% The arithmetic operations in DP alignment mainly involve comparison, addition, and min operations. Among these operations, add and min consume the most instructions. We improve the instruction efficiency and reduce required instructions by proposing the fused add-min arithmetic optimization. This is based on the observation that most add and min instructions in Listing~\ref{code:vec_banded_dp} are consecutively issued. Therefore, we merge the consecutive two add and min instruction into one fused add-min instruction (\texttt{vaddminu.vv} or \texttt{vaddminu.vx}) to reduce the required instruction number. The ISA and implementation details are given in Section~\ref{sec:isa_microarch}.

%%%%%%%%%%%%%%%%%%%%%%%%%%%%%%%%%%%%%%%%%%%%%%%%%%%%%%%%%%%%%%%%%%%%%%%%%%%%%%%%%%%%%%%%%%
\subsection{ISA Extension and Implementation}\label{subsec:isa_ext}

\begin{table}[t]
  \centering
  \footnotesize
  \caption{\design ISA extensions}
  \label{tab:isa_ext}
  \resizebox{0.95\linewidth}{!}{
  % \begin{tabularx}{\linewidth}{llX}
  \begin{tabular}{llc}
      \hline
      \textbf{Instruction} & \textbf{funct3} & \textbf{Description}  \\
      \hline
      \hline
      \texttt{vslide1band.vx vd, vs2, rs1} & \texttt{OPMVX} & Band sliding and string loading\\
      \hline 
      \texttt{vrevband.vx vd, vs2, rs1}    & \texttt{OPMVX} & String band reverse \\
      \hline 
      \texttt{vaddminu.vx vd, vs2, rs1} & \texttt{OPIVX} & \multirow{2}{*}{Fused Add-Min operation}\\
      \texttt{vaddminu.vv vd, vs2, vs1} & \texttt{OPIVV} & \\
      \hline
  \end{tabular}}
\end{table}

The dataflow optimization strategies in Section~\ref{subsec:proposed_dataflow_opt} require the support from ISA and microarchitecture. Table~\ref{tab:isa_ext} summarizes \design's specialized ISA extensions to address the performance bottlenecks identified in our profiling analysis. These custom RVV instructions target the specific memory access patterns and computational characteristics of genomic alignments, addressing the low SIMD lane utilization observed in current RVV implementations while maintaining the programmability advantages of the RISC-V architecture.

% The functionalities of the proposed ISAs include:
% \begin{enumerate}[leftmargin=*]
%     \item \texttt{vslide1band.vx} instruction operates with the \texttt{OPMVX} format and has two critical functions: 1. fine-grained band sliding and 2. fast string loading operations. These functions provide full support for the inter-sequence parallelism in Section~\ref{subsec:inter_seq_par}. The overhead of string loading can be significantly reduced using these functions. 
%     \item \texttt{vrevband.vx} with \texttt{OPMVX} format provides low-latency band reversal functionality, addressing the expensive overhead due to reference string reverse operations.
%     \item \texttt{vaddminu.vx} and \texttt{vaddminu.vv} implement the vector-scalar and vector-vector Fused Add-Min operations, respectively. This is to support the proposed Add-Min arithmetic optimization in Section~\ref{subsec:fused_addmin}, thereby reducing the required number of issued instructions.
% \end{enumerate}

\begin{figure}[t]
    \centering
    \includegraphics[width=0.95\linewidth]{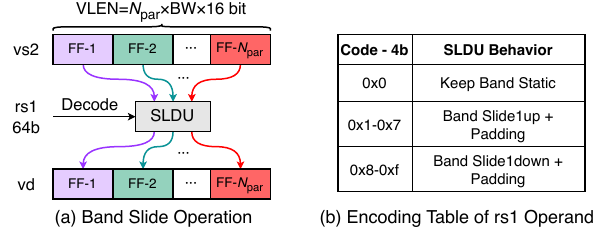}
    \caption{(a) Band slide operation using \texttt{vslide1band.vx}. (b) SLDU behavior encoding table.}
    \label{fig:reverse_bandslide}
\end{figure}

The proposed \design extensions belong to the RVV \texttt{OPMVX}, \texttt{OPIVX}, and \texttt{OPIVV} categories, operating on vector–vector or vector–scalar inputs. We enhance Ara2 with \design-specific functional units, including optimized SLDU and SIMD ALUs, to support these new ISAs for genomic computation.

\noindent\textbf{Band Slide Operation.} 
In the FFS dataflow, each forefront on the SIMD lane must shift by one element or remain static, depending on its direction. This creates a challenge for realizing inter-sequence parallelism: each band on the SIMD lane needs be independently controlled (shift or stay static). 
However, RVV 1.0~\cite{rvv_spec} lacks such fine-grained sliding, as its native \texttt{vslide1up/down} instructions act on the entire \texttt{VLEN}-bit lane. To address this limitation, we introduce the \texttt{vslide1band.vx} instruction, which enables per-band sliding and efficient string loading. \texttt{vslide1band.vx} instruction operates with the \texttt{OPMVX} format and provides two critical functions: fine-grained band sliding and fast string loading operations. 
We modify Ara2's SLDU to realize the band sliding operation. Figure~\ref{fig:reverse_bandslide}-(a) depicts how to realize band slide operation using minimal hardware cost. The 64-b \texttt{rs1} operand is decoded in SLDU, where every 4-b segment controls the corresponding band behavior using the encoding table in Figure~\ref{fig:reverse_bandslide}-(b). For example, \texttt{0x0} keeps the band static while \texttt{0x1} to \texttt{0x7} shift to higher bits by 1 element with padding. This allows up to 16 bands to be independently shifted or held with a single instruction.

\noindent\textbf{Fast String Loading and Reverse.}
For fast string loading, we reuse the same encoding table to transfer new characters of input sequences. Since DNA sequences (AGCT) can be encoded using 2 bits, \texttt{vslide1band.vx} reduces latency and memory bandwidth by transmitting only one character per band rather than loading the entire BW-length string.
\texttt{vrevband.vx} with \texttt{OPMVX} format provides low-latency band reversal functionality, addressing the expensive overhead of reference string reverse operations. We add a 256-b shuffle network to realize string reverse operation, with control signals encoded into the \texttt{rs1} operand to determine reverse width. \design supports BW from 16 to 256. Compared to RVV's \texttt{vrgather.vv} instruction, \texttt{vrevband.vx} achieves significantly lower latency.

\noindent\textbf{Fused Add-Min Implementation.} To implement the Add-Min arithmetic optimization in Section~\ref{subsec:proposed_dataflow_opt}. We develop \texttt{vaddminu.vx} and \texttt{vaddminu.vv} instructions to implement vector-scalar and vector-vector Fused Add-Min operations, respectively. To improve \design's SIMD lane utilization, we add four add-min arithmetic units to each Ara2 lane's ALU, reducing the required number of issued instructions.

%%%%%%%%%%%%%%%%%%%%%%%%%%%%%%%%%%%%%%%%%%%%%%%%%%%%%%%%%%%%%%%%%%%%%%%%%%%%%%%%%%%%%%%%%%
%%%%%%%%%%%%%%%%%%%%%%%%%%%%%%%%%%%%%%%%%%%%%%%%%%%%%%%%%%%%%%%%%%%%%%%%%%%%%%%%%%%%%%%%%%
\section{Evaluation}\label{sec:eval}

%%%%%%%%%%%%%%%%%%%%%%%%%%%%%%%%%%%%%%%%%%%%%%%%%%%%%%%%%%%%%%%%%%%%%%%%%%%%%%%%%%%%%%%%%%
\subsection{Methodology}

% \noindent\textbf{Software Implementation.}
% The AVX2 and RVV kernels for banded DP alignment are implemented using C language and compiled using the LLVM toolchain~\cite{lattner2004llvm} (version 16.0.6) with \texttt{-O3} optimization for both architectures. All kernels use 16-bit integer for computation. We perform functional verification and behavioral simulation using the RISC-V ISA simulator, SPIKE~\cite{spike}, for all RVV implementations. 
\noindent\textbf{Software Implementation.}
AVX2 and RVV kernels for banded DP alignment are implemented in C and compiled with LLVM (v16.0.6) using \texttt{-O3} optimization and 16-bit integer arithmetic. RVV implementations are functionally verified with the SPIKE RISC-V simulator~\cite{spike}.

\noindent\textbf{Hardware Simulation.}
\design is built on the Ara2 RVV processor~\cite{perotti2024ara2}, featuring a single-issue, in-order, 6-stage CVA6 core~\cite{zaruba2019cva6} and a vector unit with Zve64x and \design extensions (VLEN=4096b and 8 lanes). The 16-core system connects cores via Constellation~\cite{zhao2022constellation} and an 8-channel DDR4-3200 memory. Designs are synthesized using GF 22nm FD-SOI with Synopsys Design Compiler at 1.25 GHz. Timing and energy parameters of SRAMs are modeled using CACTI-3DD~\cite{CATCI-3DD}.

\noindent\textbf{Baselines.}
We evaluate four hardware baselines (CPU, GPU, RVV, and ASIC) by measuring the performance of DP kernels executed under bare-metal mode. We implement CPU-AVX2 kernel~\cite{marco2021wfa} to achieve throughput comparable to BSAlign \cite{shao2024bsalign} on an 8-core 7 nm AMD Ryzen 7 9700X (65 W TDP and 71 mm$^2$ die) running at 3.8–5.5 GHz with 64 GB DDR5-4800 memory. We adopt AGAThA~\cite{park2024agatha} as the GPU baseline that runs on an NVIDIA RTX 4090 card (5 nm, 450 W TDP, 609 mm$^2$ die) at 2.5 GHz with 24 GB GDDR6X memory. The Ara2-RVV baseline executes the vectorized kernel on the Ara2 processor \cite{perotti2024ara2}. Finally, ASIC baselines include programmable and RISC-V-based accelerators such as Darwin \cite{turakhia2018darwin}, GenDP \cite{gu2023gendp}, GMX \cite{doblas2023gmx}, and QUETZAL \cite{pavon2024quetzal} (Fujitsu A64FX \cite{fujitsu_a64fx}, ARM SVE2 \cite{sve2_spec}).

% \begin{table}[t]
% \scriptsize
% \centering
% \caption{Baseline CPU and GPU configurations}
% \label{tab:ryzen7}
% \begin{tabular}{c|c}%{0.8\linewidth}{p{2.3cm} X}
% \hline
% \textbf{CPU}                      & AMD Ryzen 7 9700X with 8 cores \\ 
% \hline
% \textbf{Cache Size}               & 640 KB (L1) 8 MB (L2) 32 MB (L3) \\ 
% \hline
% \textbf{Clock Frequency}          & 3.8-5.5 GHz                    \\ 
% \hline
% \textbf{Technology}               & 7nm                            \\ 
% \hline
% \textbf{TDP}                      & 65W                       \\ 
% \hline
% \textbf{Die Area}                 & 71 mm$^2$                       \\ 
% \hline
% \textbf{Memory}                   & 2-channel 64GB DDR5-4800     \\ 
% \hline
% \hline
% \textbf{GPU}                      & NVIDIA RTX4090 \\ 
% \hline
% \textbf{Cache Size}               & 128 KB (L1) 72 MB (L2) \\ 
% \hline
% \textbf{Boost Frequency}          & 2.5 GHz                    \\ 
% \hline
% \textbf{TDP}                      & 450W                       \\ 
% \hline
% \textbf{Technology}               & 5nm                            \\ 
% \hline
% \textbf{Die Area}                 & 609mm$^2$                       \\ 
% \hline
% \textbf{Memory}                   & 24GB GDDR6X, 1.01 TB/s     \\ 
% \hline
% \end{tabular}
% \end{table}

%%%%%%%%%%%%%%%%%%%%%%%%%%%%%%%%%%%%%%%%%%%%%%%%%%%%%%%%%%%%%%%%%%%%%%%%%%%%%%%%%%%%%%%%%%
% \subsection{Baselines and Workloads}

\noindent\textbf{Baselines.}
We evaluate four hardware baselines (CPU, GPU, RVV, and ASIC) by measuring the performance of DP kernels executed under bare-metal mode. We implement CPU-AVX2 kernel~\cite{marco2021wfa} to achieve throughput comparable to BSAlign \cite{shao2024bsalign} on an 8-core 7 nm AMD Ryzen 7 9700X (65 W TDP and 71 mm$^2$ die) running at 3.8–5.5 GHz with 64 GB DDR5-4800 memory. We adopt AGAThA~\cite{park2024agatha} as the GPU baseline that runs on an NVIDIA RTX 4090 card (5 nm, 450 W TDP, 609 mm$^2$ die) at 2.5 GHz with 24 GB GDDR6X memory. The Ara2-RVV baseline executes the vectorized kernel on the Ara2 processor \cite{perotti2024ara2}. Finally, ASIC baselines include programmable and RISC-V-based accelerators such as Darwin \cite{turakhia2018darwin}, GenDP \cite{gu2023gendp}, GMX \cite{doblas2023gmx}, and QUETZAL \cite{pavon2024quetzal} (Fujitsu A64FX \cite{fujitsu_a64fx}, ARM SVE2 \cite{sve2_spec}).

% We evaluate four hardware baselines (CPU, GPU, RVV, and ASIC) by measuring DP kernel performance. RVV kernels run in bare-metal mode. The CPU baseline is an 8-core 7 nm AMD Ryzen 7 9700X (65 W TDP, 71 mm$^2$ die) operating at 3.8–5.5 GHz with 64 GB DDR5-4800 memory. The GPU baseline uses an NVIDIA RTX 4090 (5 nm, 450 W TDP, 609 mm$^2$ die) with a 2.5 GHz boost clock and 24 GB GDDR6X memory.
% \begin{itemize}[leftmargin=*]
%   \item \textbf{CPU Baselines}: We develop \textbf{CPU-AVX2} based on the benchmark code in~\mbox{\cite{marco2021wfa}} to implement Listing~\mbox{\ref{code:vec_banded_dp}}. Our developed kernel achieves comparable throughput as BSAlign~\mbox{\cite{shao2024bsalign}}.
%   \item \textbf{GPU}: \textbf{AGAThA}~\mbox{\cite{park2024agatha}} optimized for long-read alignment is adopted as the GPU baseline.
%   \item \textbf{Ara2-RVV}: We develop the vectorized kernel in Listing~\mbox{\ref{code:rvv_vec_banded_dp}} using RVV instructions and deploy on the original Ara2~\mbox{\cite{perotti2024ara2}}, which is fully compatible with RVV 1.0 specifications~\mbox{\cite{rvv_spec}}.
%   \item \textbf{ASIC Baselines}: We consider a wide variety of programmable ASIC designs, including \textbf{Darwin}~\mbox{\cite{turakhia2018darwin}}, \textbf{GenDP}~\mbox{\cite{gu2023gendp}} based on customized ISA, \textbf{GMX}~\mbox{\cite{doblas2023gmx}} based on RISC-V scalar ISA, and \textbf{QUETZAL}~\mbox{\cite{pavon2024quetzal}} based on Fujitsu A64FX SoC~\mbox{\cite{fujitsu_a64fx}} and ARM SVE2~\mbox{\cite{sve2_spec}}.
% \end{itemize}

\noindent\textbf{Workloads.}
We evaluate the two widely used workloads in genomic analysis, \textit{Edit Distance} and \textit{Smith-Waterman-Gotoh (SWG)} algorithms. SWG uses affine penalties: $match=0$, $mismatch=4$, $o=6$ $e=2$. We evaluate four lengths, including two short-read datasets (100bp and 200bp) and two long-read datasets (1kbp and 10kbp). 
Real-world datasets used in~\mbox{\cite{liu2023block,marco2021wfa,shao2024bsalign}} are adopted as the evaluation datasets. 

\subsection{Area and Power Analysis}
% Table~\ref{tab:ppa_results} presents the synthesized area and power results of \design. The original design comprising the CVA6 core~\cite{zaruba2019cva6} with Ara2~\cite{perotti2024ara2} RVV processor takes 0.796 mm$^2$ of area and 65.3 mW of power consumption. In comparison, our proposed \design core implementation with customized ISA extensions achieves a slightly larger area of 0.804 mm$^2$ with a power consumption of 67.3 mW. 
% Only 1.0\% and 3.1\% of area and power overhead are required. The overhead mainly comes from the added shuffle network in SLDU and the fused add-min ALUs in each Lane. We also use \design to extend to the 16-core \design processor, which consumes 12.898 mm$^2$ area and 1090.4 mW power dissipation. The complete \design multi-core design integrates specialized functional units that significantly accelerate genome sequencing operations while maintaining reasonable area and power costs.

Table~\ref{tab:ppa_results} summarizes the synthesized area and power results of \design. Our \design core with customized ISA extensions slightly increases the area and power of baseline CVA6~\cite{zaruba2019cva6}+Ara2~\cite{perotti2024ara2} design by only 1.0\% and 3.1\% overhead, mainly from the added shuffle network and fused add–min ALUs. The 16-core \design processor consumes 12.898 mm$^2$ and 1090.4 mW, delivering substantial acceleration for genome alignment with minimal area and power cost.

\begin{table}[t]
  \footnotesize
  \centering
  \caption{\design area and power results}
  \label{tab:ppa_results}
  \renewcommand{\arraystretch}{1.1}
  \begin{tabularx}{0.85\linewidth}{l|l|r|r}
      \hline
      \multicolumn{2}{l|}{\textbf{Design/Components}} & \textbf{Area (mm$^2$)} & \textbf{Power (mW)} \\
      \hline
      \hline
      \multicolumn{2}{l|}{CVA6 Core} & 0.125 & 17.3 \\
      \hline
      \hline
      \multicolumn{2}{l|}{Original Ara2} & 0.670 & 48.0 \\
      \hline
      \hline
      \multirow{5}{*}{\rotatebox[origin=c]{90}{\design}} & Ara2 Lane $\times8$ & 0.518 & 58.6 \\
      & Ara2 SLDU  & 0.022 & 1.6 \\
      & Ara2 VLSU  & 0.053 & 3.0 \\
      & Ara2 MASKU & 0.052 & 2.4 \\
      & Others    & 0.006 & 0.7 \\
      \hline
      \hline
      \multicolumn{2}{l|}{Original  CVA6+Ara2}      & 0.796 &  65.3 \\
      \multicolumn{2}{l|}{\design Core (CVA6+Ara2)} & 0.804 &  67.3 \\
      \hline
      \multicolumn{2}{l|}{\design On-chip Bus}   & 0.218  & 14.1 \\
      \multicolumn{2}{l|}{\design 16-core Processor}& 12.898 &  1,090.4 \\
      \hline
  \end{tabularx}
\end{table}

%%%%%%%%%%%%%%%%%%%%%%%%%%%%%%%%%%%%%%%%%%%%%%%%%%%%%%%%%%%%%%%%%%%%%%%%%%%%%%%%%%%%%%%%%%
\subsection{Performance Comparison}

\begin{figure}[t]
    \centering
    \begin{subfigure}[b]{\linewidth}
        \centering
        \includegraphics[width=\linewidth]{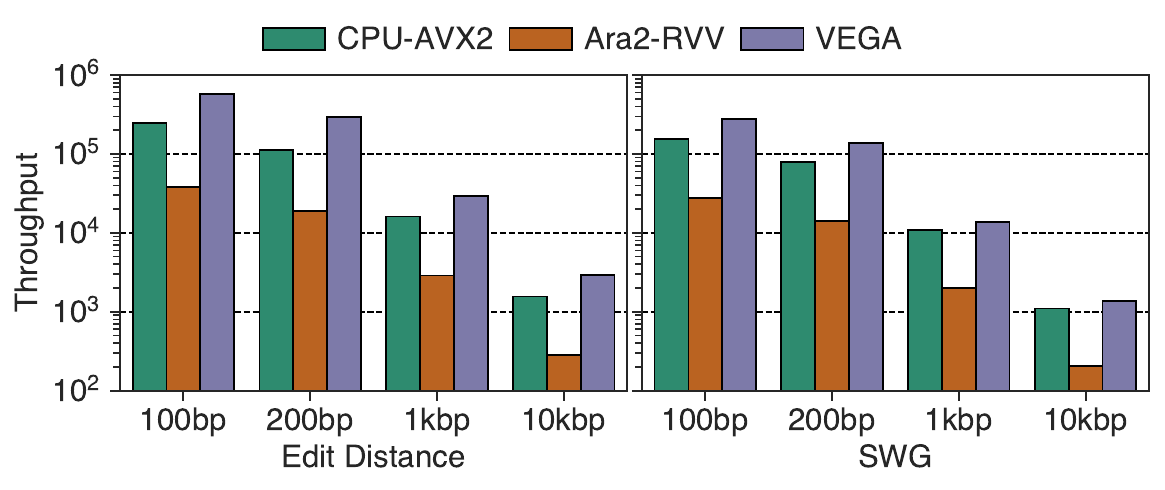}
        \caption{Single-core performance.}
        \label{subfig:perf_single_core_vs_rvv_avx2}
    \end{subfigure}   
    \begin{subfigure}[b]{\linewidth}
        \centering
        \includegraphics[width=\linewidth]{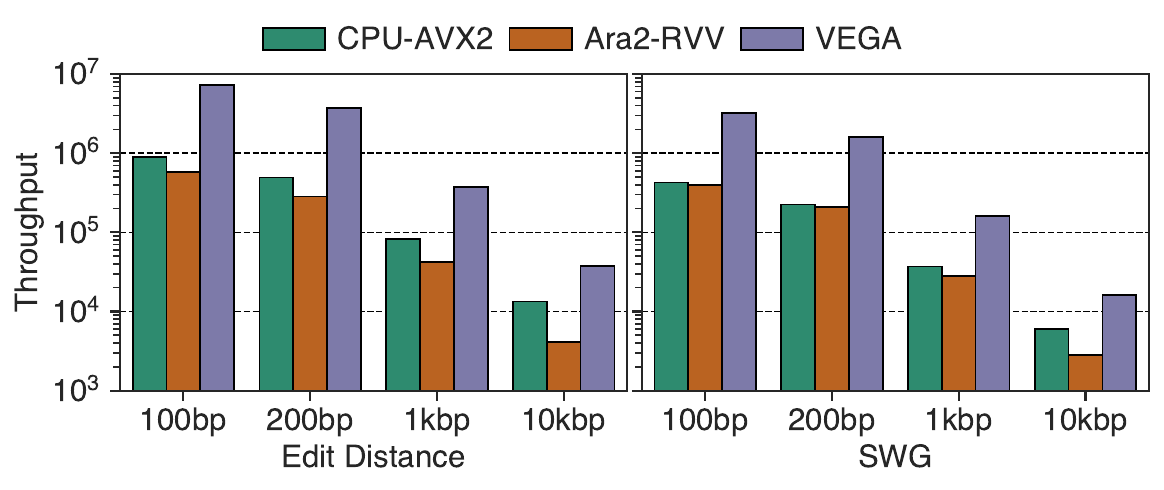}
        \caption{Multi-core performance.}
        \label{subfig:perf_multi_core_vs_rvv_avx2}
    \end{subfigure}   
    \caption{Throughput comparison with AVX2 and RVV baselines.}
    \label{fig:perf_vs_rvv_avx2}
\end{figure}

\noindent\textbf{Single- and Multi-core Performance.}
Fig.~\ref{fig:perf_vs_rvv_avx2}-(a) compares \design's single-core throughput with CPU-AVX2 and Ara2-RVV baselines. For short reads, \design delivers up to $15.1\times$ and $9.7\times$ gains over Ara2-RVV for Edit Distance and SWG, and still achieves $10.2\times$ and $6.7\times$ speedups for long 10kbp sequences. Its smaller advantage over CPU-AVX2 arises from the latter's higher frequency (5.5 GHz vs. 1.25 GHz) and larger cache hierarchy. Despite these factors, \design sustains practical throughput—2,946 alignments/s (Edit Distance) and 1,361 alignments/s (SWG) for 10kbp reads. A 16-core \design with 8-channel DDR4-3200 memory (Fig.~\ref{fig:perf_vs_rvv_avx2}-(b)) achieves $12.7\times$ and $8.0\times$ higher throughput than Ara2-RVV for short reads, and maintains $5.6\times$–$8.9\times$ gains for 10kbp sequences, demonstrating promising scalability and parallel efficiency.

\begin{table}[t]
  \centering
  \caption{Area and power efficiency comparison of long-read banded SWG designs.}
  \label{tab:comp_cpu_gpu_eff}
  % \scriptsize
  \footnotesize
  \resizebox{\linewidth}{!}{
  \begin{threeparttable}
      \renewcommand{\arraystretch}{1.1}
      \setlength{\tabcolsep}{2.5pt}
          % \begin{tabularx}{\linewidth}{c|ccccc}
        \begin{tabular}{c|ccccc}
            \hline
            \textbf{Design} & CPU-AVX2  & QUETZAL~\cite{pavon2024quetzal} & GMX-SWG~\cite{doblas2023gmx} & AGAThA~\cite{park2024agatha} & \design (Ours) \\
            \hline
            \hline
            \textbf{Platform} & CPU & CPU & CPU+ASIC & GPU & VPU \\
            \hline
            \multirow{2}{*}{\textbf{ISA}} & x86 & ARM & Customized & \multirow{2}{*}{PTX} & RISC-V \\
                          & AVX2 & SVE2 & RV64G & & RV64GCV \\
            \hline
            \textbf{VLEN} & 256b & 512b & / & / & 4096b \\
            \hline
            \textbf{Area (mm$^2$)}  & 71 & 2.9 & 1.2 & 609 & 12.9 \\
            \hline
            % \textbf{Power (mW)} & 65,000 & / & 397.4 &  & 1090.4 \\
            % \hline
            \hline
            \multirow{2}{*}{\textbf{Tp./mm$^2$ $\uparrow$}\tnote{*}} & 4.9 & 3.7\tnote{\textdagger} & 1,625.0 & 40.6 & 1,241.5 \\
             & (1.3$\times$) & (1.0$\times$) & (439.2$\times$) & (11.0$\times$) & (335.5$\times$) \\
            \hline
            \multirow{2}{*}{\textbf{Align./mJ $\uparrow$}\tnote{*}} & 0.5 & \multirow{2}{*}{/} & 5.1 & 1.8 & 14.7 \\
            & (1.0$\times$) &  & (10.2$\times$) & (3.6$\times$) & (29.4$\times$) \\
            \hline
        \end{tabular}
        \begin{tablenotes}
           % \item [*] Values for 10kbp reads. 
           \item [*] Scaled to 22nm and 10kbp. 
           $^\dagger$ Scaled from 10kbp and 30kbp to 10kbp.
        \end{tablenotes}
  \end{threeparttable}
  }
\end{table}

\noindent\textbf{Comparison with Existing Accelerators.}
Table~\ref{tab:comp_cpu_gpu_eff} compares \design to state-of-the-art CPU and GPU baselines~\cite{pavon2024quetzal,park2024agatha}. All designs are scaled to 22nm using methodologies in~\cite{sarangi2021deepscaletool} for fair comparison. \design achieves an area efficiency of 1,241.5 throughput/mm$^2$---the highest among programmable designs and second only to the hard-wired GMX-SWG ASIC---delivering $253.4\times$ and $30.6\times$ improvements over CPU-AVX2 and GPU baselines, respectively. \design also attains the highest energy efficiency of 14.7 alignments/mJ among all compared designs. The superior performance gains stem from the specialized vector extensions and large 4096-bit vector length in \design.
Notably, QUETZAL represents the other genomic accelerator based on ARM SVE ISA. It is mainly optimized for reduced-latency memory access, thereby providing speedup. However, QUETZAL shows inferior long-read performance compared to \design because it adopts a fixed 512b vector length that fails to exploit inter-sequence parallelism. 
In contrast, \design with flexible vector length effectively exploits the data parallelism of various levels to provide higher throughput and efficiency.

%%%%%%%%%%%%%%%%%%%%%%%%%%%%%%%%%%%%%%%%%%%%%%%%%%%%%%%%%%%%%%%%%%%%%%%%%%%%%%%%%%%%%%%%%%
% \subsection{Sensitivity Analysis}

\begin{figure}[t]
  \centering
  \includegraphics[width=\linewidth]{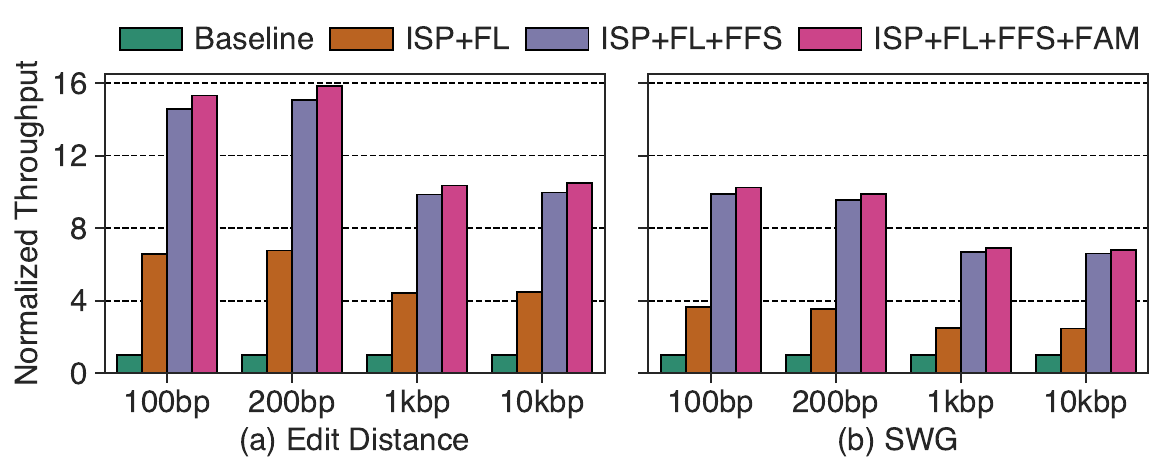}
  \caption{Incremental improvements of \design optimizations (ISP=inter-seq. parallelism, FL=fast string load, FFS=forefront stationary dataflow, and FAM=fused add-min).}
  \label{fig:perf_incr_improvement}
\end{figure}

\noindent\textbf{Impact of Optimizations.}
Fig.~\ref{fig:perf_incr_improvement} shows the incremental contributions of the proposed \design optimizations over the Ara2-RVV baseline across different sequence lengths and DP algorithms. 
The forefront stationary dataflow (FFS) and fast string loading (FL) provide the two most substantial performance gains. The fused add-min (FAM) operations provide additional 10\% throughput improvement due to the lower instruction overhead. The full optimization stack (ISP+FL+FFS+FAM) delivers up to $15\times$ throughput improvement over the Ara2-RVV baseline. The consistent performance advantages across sequence lengths and algorithm types.

\section{Conclusion}
In this paper, we have presented \design, a new and scalable software-hardware co-design that accelerates genomic alignment through specialized RVV extensions. Our approach has introduced three key innovations: Forefront Stationary dataflow that minimizes memory bandwidth; inter-sequence parallelism that maximizes SIMD lane utilization; and specialized vector instructions with minimal area overhead. \design effectively balances efficiency, programmability, and scalability—delivering significant performance improvements while maintaining algorithmic flexibility.
% We present \design, a scalable RVV-based co-design for genomic alignment. It features a Forefront Stationary dataflow, inter-sequence parallelism, and specialized vector instructions, balancing efficiency, programmability, and scalability with significant performance gains.

%%%%%%%%%%%%%%%%%%%%%%%%%%%%%%%%%%%%%%%%%%%%%%%%%%%%%%%%%%%%%%%%%%%%%%%%%%%%%%%%%%%%%%%%%%
% \section*{Acknowledgements}
% This work is partially supported by the SRC JUMP 2.0 PRISM Center, and National Science Foundation (NSF) grants \#1826967, \#1911095, \#2052809, \#2112665, \#2112167, and \#2100237.

\clearpage
\begin{footnotesize}
    \bibliographystyle{IEEEtran}
    \bibliography{refs.bib}
\end{footnotesize}

\end{document}